# A Parallel Genetic Algorithm for Three Dimensional Bin Packing with Heterogeneous Bins

Drona Pratap Chandu

*Amazon Development Centre India*
*Chennai, India*

*Abstract*— This paper presents a parallel genetic algorithm for three dimensional bin packing with heterogeneous bins using Hadoop Map-Reduce framework. The most common three dimensional bin packing problem which packs given set of boxes into minimum number of equal sized bins is proven to be NP Hard. The variation of three dimensional bin packing problem that allows heterogeneous bin sizes and rotation of boxes is computationally more harder than common three dimensional bin packing problem. The proposed Map-Reduce implementation helps to run the genetic algorithm for three dimensional bin packing with heterogeneous bins on multiple machines parallely and computes the solution in relatively short time.

## I. Introduction

The three dimensional bin packing problems is to pack the given three dimensional boxes into a subset of given set of containers such that solution maximises the ratio of space occupied by given boxes to the total capacity of chosen subset of containers. There are many variants of one dimensional bin packing ,two dimensional bin packing and three dimensional bin packing. The common three dimensional bin packing problem(3D-BPP) considers equal sized bins and don't allows rotation of the boxes. The bin packing has many applications such as loading cargo into vehicles. In this paper we consider three dimensional bin packing with heterogeneous bins where bin sizes can be different and also allows the rotation of boxes to be packed such that a box can be placed into the one in any of six possible orientations and this is computationally more difficult than original three dimensional bin packing.

Three dimensional bin packing problem (3D-BPP) is a NP-Hard problem as it generalise one dimensional bin packing problem that was proven to be NP-Hard. The running time of three dimensional bin packing is long and parallel computing is useful to compute an exact or near optimal solution to three dimensional bin packing in a reasonable time. This paper presents a parallel genetic algorithm for three dimensional bin packing with heterogeneous bins using Hadoop Map-Reduce framework. The proposed Map-Reduce implementation helps to run the genetic algorithm for three dimensional bin packing on multiple machines parallely and computes the solution in relatively short time.

## II. Related work

Many algorithms were proposed for three dimensional bin packing with equal sized bins(3D-BPP). Chen[4] provides a mixed integer programming formulation to solve the 3D-BPP without orientation restriction. Martello and Den[5] present an exact algorithm for filling a single bin. Martello[6] present an algorithm to solve moderately large instances optimally for the general 3D-BPP and its robot-packable variant. George and Robinson[7] present a wall-building approach heuristic without restrictions on the box orientation for identical bins.

Bischoff and Janetz[8] propose a pallet loading heuristic for non-identical bins where the stability of the pallet is considered. Pisinger[9] offers a heuristic based on the wall-building approach, where strips and layers are created so that the 3D-BPP can be decomposed into smaller sub-problems. Baltacio glu and Moore[10] present a new heuristic algorithm using rules to mimic human intelligence to solve the 3D-BPP. Faroe[11] provides a heuristic for packing boxes into a minimum number of identical bins based on guided local search without box rotation is allowed.

Gehring and Bortfeldt[12] present a genetic algorithm for loading strongly heterogeneous sets of boxes into a single bin. Hopper and Turton[13] used genetic algorithms to solve the 3D-BPP. Zhang[14] propose a heuristic algorithm to solve the generic 3D-BPP. Crainic[15] introduce a two-level tabu search where the first level aims to reduce the number of bins and the second optimizes the packing of the bins for boxes with fixed orientations. Lodi[16] provide a tabu search framework by exploiting a constructive heuristic to evaluate the neighbourhood where the box orientation is fixed.

Xeuping[1] propose a hybrid genetic algorithm for three dimensional bin packing problem with heterogeneous bins and rotation of boxes.

## III. Hadoop map-reduce

Map-Reduce is a programming model developed by Dean and Ghemawat[2] for writing highly scalable parallel programs. User of Map-Reduce framework have to implement two functions map and reduce. Scheduling, distributing, aggregation of the results and failure recovery was automatically handled by Map-Reduce runtime system. Apache is providing an open source implementation of Map-Reduce as a part of its Hadoop project.

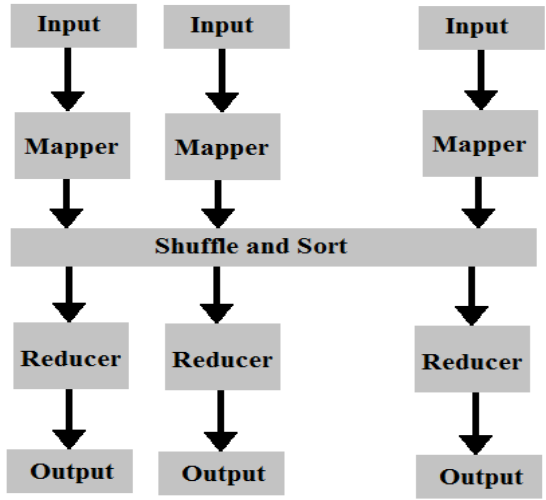

Figure 1 . Hadoop Map-Reduce

The map method takes a key value pair as input and outputs key value pairs. The output key value pairs of map functions running on multiple mapper machines is collected by runtime system and it sort them and group them according to key value. Then it passes a key and list of values corresponding to that key to reduce function. The reduction process the given (key ,values list) and outputs (key ,value) pair. Map operations and reduce operations are distributed among several machines in Hadoop cluster as shown in Figure 1. Many real world problems can be solved using Map-Reduce model.

## IV. A GENETIC ALGORITHM FOR THREE DIMENSIONAL BIN PAKCING

Genetic algorithms(GA's) have been applied to many optimization problems. GA's are proven to be efficient in finding globally optimal solutions. GA's are computational models inspired by evolutionary theory. A genetic algorithm(GA) begins with a random sample of chromosomes where each chromosome represents a solution to the given problem. GA selects set of chromosomes from current generation to be used as mating pool and applies operators like mutation,crossover on the selected chromosomes to create next generation. These operators mutation, crossover and selection resemble a process of natural evolution.Creation of new generation from previous generation is repeated until some convergence condition is satisfied.

Xueping[1] proposed a genetic algorithm for three dimensional bin packing with heterogeneous bins and rotation of boxes. Genetic algorithm frame works used by Xueping[1].

Genetic Algorithm proposed by Xueping[1] has the following steps

1. Create $Z$ random chromosomes including four special chromosomes.
2. Compute fitness of chromosomes in current generation.
3. Select $E$ chromosomes with high fitness and add them to next generation.
4. Select $Z$-$E$ pairs of chromosomes and apply cross over , mutation operations on selected chromosomes to produce a pair of new chromosomes in the next generation.
5. Repeat steps 2 to 4 for desired number of times.

The elements of proposed genetic algorithm are described below.

### I. chromosome Encoding

The chromosome used by Xueping[1] consists of two parts box packing sequence(BPS) $S^1$ and container loading sequence (CLS) $S^2$. Box packing sequence is a permutation of {1,2,.....,M} where M is number of boxes. Container loading sequence is a permutation of {1,2.......,N} where N is number of bins.

### II. Population Initialisation

Four special chromosomes whose box packing sequence is sorted by volume,length , width and height with randomly generated container loading sequence are added to initial population. Rest of the chromosomes are generated randomly.

### III. selection

Xueping[1] used tournament method for selection. The fitness of each chromosome is computed by heuristic packing strategy described in section V. Chromosomes are sorted in descending order of fitness. First E chromosomes are selected as elitism which directly proceed into the next generation. (Z - E) parents are selected into the mating pool using the tournament selection method with the tournament size equal to 2. In each round, it selects two different chromosomes from population, with probability based on fitness the better one(larger fitness) is added into the mating pool, otherwise the weak one is added to mating pool. It selects successive two chromosomes in the mating pool as parents. For each pair of parents, there is a probability *probc* that they go directly into next generation, otherwise two childs are generated through crossover.

### IV. Crossover and Mutation

Xueping[1] used crossover operation with two cut points suitable for order based encoding. The two parts s1 and s2 of chromosome are computed independently during crossover operation. Two cutting points are randomly selected for the gene sequence, say i and j. Parents P1 and P2 will generate two child chromosomes O1 and O2. Child O1's genes sequence s1 are generated as follows, genes at positions from i+1 to j are copied from i+1 to j in P1. Other positions in O1 starting from j+1 are filled circularly by genes starting from j+1 in P2 which are not already in O1.

Child O2 is obtained by exchanging the roles of P1 and P2. Mutation is conducted on each newly generated chromosomes O1 and O2 with probability *Pm*. For each gene sequence, two positions are randomly selected and genes on these positions are swapped.

## V. BEST MATCHING HEURISTIC PACKING STRATEGY

Xueping[1] used the concept of empty maximal spaces (EMS's) to represent list of largest empty cubic spaces. An empty container is represented by single empty maximal space. When one box is place on corner of empty container , it creates three maximal empty spaces. Each of the three is a rectangular shape that is not overlapping with box such that two of its dimensions are equal to corresponding dimensions of empty box and third is equal to difference of third dimension of box and corresponding dimension of box.

### I. *Priority of Empty Maximal Spaces*

Xueping[1] defined priority using following rules for two empty maximal spaces with their minimum vertex coordinates are (x1,y1,z1) and (x2,y2,z2).

1. The strategy compares the smallest coordinates value of the two vertexes, the vertex with smaller one will be given high priority.
2. if compared values are equal, it compares the second smallest values and gives high priority to the vertex with smaller one.
3. if the two compared values are equal again, it compares the third smallest values and assigns highest priority to vertex with smallest value.

### II. *Placement Selection*

In each iteration, first *Kb* unpacked boxes are chosen based on the order given by BPS and first *Ke* EMS's in the bin currently considered are selected with the order defined by priority. For these *Kb* boxes and *Ke* EMS's, all the feasible placements assignment with 6-way orientations of box are computed. The (box , empty maximal space) pair with the largest fill ratio is first chosen for the placement.

When one box has several feasible placements in one empty space, the one has smallest margin is selected. This can be done by calculating as follows: the dimensions of the space is (X1 – x1,Y1 – y1,Z1 - z1) as defined, define the dimensions of the boxes after rotation as (l¹ ,w¹ ,h¹ ), substract two vectors and we get (X1 – x1 – l¹ ;Y1 - y1 – w¹ ;Z1 - z1 – h¹ ) which represents the margins to three faces of the space. Then, the priority of the placement is defined the same as the one in selecting empty maximal space.If there is no feasible placement assignment for current boxes and spaces selection, we will try next ke EMS's in the container, the process continues until at least one feasible placement is find. If all EMS's in current container are tried, we move forward to try EMS's in the next opened container. If all opened containers are tried without finding one feasible placement, we open the next unopened container in order given by CLS.

*OC = φ*  // Let OC be the list of opened containers
while *BPS != φ* do
{ //Let *P* be a priority queue of candidate placements
  boxplaced ← false
  for each opened container *c* in *OC* do
  {  // Let EMS's be the empty maximal spaces in c
    *j* ← 1
    while *j* <= EMS's:size() and boxplaced = false do
    { *k* ← *j+ke*
      while *j* < *k* and *j* <=EMS's:size() do
      {for all 6 orientations *bo* do
        {if Box *BPSi* can be placed in EMS's *j* with
              orientation *bo* then
          {Add this placement combination to *P*
          }
        }
      }
      *j* ← *j+1*
    }
    if *P !=*  *φ* then
    {Make the placement indicted by *P1*
      Update EMS's
      boxplaced ← true
    }
  }
  if boxplaced = true then
  { break
  }
}
while *CLS != 0* and boxplaced = false do
{ Let *EMS* be the initial empty space in *CLS1*
  *OC ← OC U CLS1*
  *CLS ← CLS\CLS1*
  for *i = 1* to *kb* and *i != BPS:size()* do
  {for all 6 orientations *bo* do
    {if Box *BPSi* can be placed in EMS with orientation *bo*
                  then
      {Add this placement combination to *P*
      }
    }
  }
  if *P != φ* then
  { Make the placement indicted by *P1*
    Update EMS's
    boxplaced = true
  }
}
if boxplaced = false then
{ return null
}
}
return packaging solution.
**Figure 2. Best matching heuristic packing strategy**

If there is no container in the list, the algorithm stops without finding a feasible packing solution.

Chosen item is placed in the chosen empty space and the item is removed from the unpacked items list and EMS's list is updated . If there is no element in the unpacked boxes list, the algorithm terminates with solution found and the fitness of the solutions is set to be the fill ratio.

## VI. IMPLEMENTATION USING HADOOP

The proposed model requires to run the Map-Reduce iteratively. Each iteration takes population of one generation as input and applies necessary evolutionary operators during Map-Reduce to produce the next generation as output. Only one mapper is used and number of reducers can be increased to compute the solutions quickly.

Overall process can be summarised as follows
1. Create initial population randomly.
2. Compute fitness of each solution in the initial population.
3. Run Map-Reduce with (solution,fitness) as input.
4. Collect output (solution,fitness) pairs of reduce phase.
5. Repeat steps 3 to 4 with output collected in step 4.
6. Output the best solution in the last generation.

Map function receives (solution,fitness) pairs as input. During Map phase precomputed fitnesses of solutions are used to select the solutions by tournament selection. Solutions and their fitnesses are stored in buffer until all chromosomes are received by Map. Solutions are sorted by fitness and E solutions having best fitness are added to next generation without applying evolutionary operators. For these elite E chromosomes map function outputs a pair of chromosomes first one is current chromosome in the elite E and second is empty string representing illegal chromosome.

To generate the rest of the population of next generation, pairs of chromosomes are selected and emitted as output for the reduce phase. The pairs of chromosomes generated by map function are used as parent chromosomes by reduce function to produce next generation chromosomes.

During the reduce phase evolutionary operators are applied to generate child chromosomes from pair of chromosomes emitted by map phase. When reduce function receives a pair of chromosomes with second chromosome is illegal, reduce just emits the first chromosome and its fitness as output. Best Match Heuristic Packaging Procedure presented in section V is used to compute the fitness of solutions. When reduce function receives a pair of two legal chromosomes, it applies crossover operator to produce two child chromosomes. Then mutation operator is applied over two child chromosomes. The reduce emits (solution,fitness) pairs as output .Implementation of Mapper using Java programming language is shown in Figure 3.

```
package ThreeDBinPack;

import java.io.IOException;
import java.util.Arrays;
import org.apache.hadoop.io.*;
import org.apache.hadoop.mapred.*;

public class BinPack3DMapper extends MapReduceBase
        implements Mapper<Text, DoubleWritable,Text,
        Text>
{  int populationSize = 100;
  int processedItems = 0;
  double[] fitnessVector = new double[populationSize];
  String  population[] = new String[populationSize];
  Text parentOne = new Text("");
  Text parentTwo = new Text("");
  private int E;

  @Override
  public void map(Text key, DoubleWritable
        value,OutputCollector<Text, Text> outputCollector,
        Reporter reporter) throws IOException
{   population[processedItems] = key.toString();
   fitnessVector[processedItems]= value.get();
   processedItems++;
   if( processedItems == populationSize)
   {  MapHelper.sortSolDescendingByFitness
        (population,fitnessVector);
     parentTwo.set("");
     for(int i=0;i<E;i++)
     {  parentOne.set(population[i]);
        outputCollector.collect(parentOne, parentTwo);
     }
     for(int i=0;i < populationSize - E;i++)
     {   String par1 = MapHelper.selectRandom
          (population,fitnessVector);
        String par2 = MapHelper.selectRandom
          (population,fitnessVector);
       parentOne.set(par1);
       parentTwo.set(par2);
       outputCollector.collect(parentOne, parentTwo);
     }
     processedItems = 0;
     Arrays.fill(population, null);
   }
  }
}
```

**Figure 3.  Implementation of Mapper using Java**

Implementation of Reducer using Java programming language is shown in figure 4.

```
package ThreeDBinPack;

import java.io.IOException;
import java.util.Iterator;
import org.apache.hadoop.io.*;
import org.apache.hadoop.mapred.*;

public class BinPack3DReducer extends MapReduceBase
        implements Reducer<Text, Text, Text,
        DoubleWritable>
{   private Text outputKey = new Text();
    private DoubleWritable outputValue = new
        DoubleWritable();

    @Override
    public void reduce(Text solution, Iterator<Text>
        solution2,
    OutputCollector<Text, DoubleWritable> results, Reporter
        reporter) throws IOException
{   String parent1 = solution.toString();
    String parent2 = solution2.next().toString();
    String child1 = null;
    String child2 = null;
    if(parent2.length() == 0)
    {   child1 = parent1;
    } else
    {   int cutpoint = (int) (Math.random() *
        parent1.length());
      child1 = ReduceHelper.firstChildByCrossover
        (parent1,parent2,cutpoint);
      child2 = ReduceHelper.secondChildByCrossover
        (parent1,parent2,cutpoint);
      child1 = ReduceHelper.mutate(child1);
      child2 = ReduceHelper.mutate (child2);
    }
    double fitness = ReduceHelper.computeFitness (child1);
    outputKey.set(child1);
    outputValue.set(fitness);
    results.collect(outputKey, outputValue);

    if(child2 != null){
      fitness = ReduceHelper.computeFitness (child2);
      outputKey.set(child2);
      outputValue.set(fitness);
      results.collect(outputKey, outputValue);
    }
  }
}
```

**Figure 4.  Implementation of Reducer using Java**

The Map-Reduce execution is repeated until convergence condition is satisfied.

## VII.   CONCLUSIONS

This paper presents a parallel genetic algorithm for generalized vertex cover problem using Hadoop Map-Reduce framework. In this implementation fitness computation operations,crossover operations, and mutation operations are distributed among all the machines in Hadoop cluster running reduce phase. This parallel implementation provides a method to compute vertex cover in a relatively short time by running parallely on multiple machines and running time decreases as the number of machines in Hadoop cluster running reduce phase are increased.

### ACKNOWLEDGEMENT

I would like to thank all my colleagues at Amazon India Development Centre, colleagues at Freescale Semiconductor for their participation in long discussions with me and all my classmates at Indian Institute of Technology Roorkee who used to participate in discussions with me.

### REFERENCES

[1] Xueping Li, Zhaoxia Zhao, Kaike Zhang ,"A genetic algorithm for the three-dimensional bin packing problem with heterogeneous bins" in proceedings of the 2014 Industrial and Systems Engineering Research Conference.
[2] AJeffrey Dean and Sanjay Ghemawat, "Map-Reduce: Simplified Data Processing on Large Clusters" in Proceedings of the 6th USENIX Symposium on Operating Systems Design and Implementation 2004, 137-149.
[3] http://hadoop.apache.org/
[4] Chen, C. S., Lee, S. M., and Shen, Q. S., "An analytical model for the container loading problem", European Journal of Operational Research,1995, 80(1):68–76.
[5] Den Boef, E., Korst, J., Martello, S., Pisinger, D., and Vigo, D.,"Erratum to The three-dimensional bin packing problem: Robot-packable and orthogonal variants of packing problems", Operations Research, 2005,53(4):735–736.
[6] Martello, S., Pisinger, D., Vigo, D., Den Boef, E., and Korst, J., "Algorithm 864: General and robot-packable variants of the three-dimensional bin packing problem", ACM Transactions on Mathematical Software,2007, 33(1):1–12
[7] George, J. A. and Robinson, D. F. , "A heuristic for packing boxes into a container. Computers and Operations Research", 1980,7(3):147–156.
[8] Bischoff, E. E., Janetz, F., and Ratcliff, M. S. W. , "Loading pallets with non-identical items", European Journal of Operational Research, 1995, 84:681–692.
[9] Pisinger, D., "Heuristics for the container loading problem. European Journalof Operational Research", 2002,141:382–392.
[10] Baltacio glu, E., Moore, J. T., and Hill Jr., R. R., "The distributor's threedimensional pallet-packing problem: a human intelligence-based heuristic approach", International Journal of Operational Research, 2006,1(3):249–266.
[11] Faroe, O., Pisinger, D., and Zachariasen, M., "Guided local search for three-dimensional bin-packing problem" INFORMS Journal on Computing, 2003,15(3):267–283.
[12] Gehring, H. and Bortfeldt, A., "A genetic algorithm for solving the container loading problem" International Transactions in Operational Research, 1997, 4:401–418.

[13] Hopper, E. and Turton, B. C. H., "An empirical investigation of metaheuristics and heuristic algorithms for a 2D packing problem", European Journal of Operations Research, 2001, 128:34–57.

[14] Zhang, D.-F., Wei, L.-J., Chen, Q.-S., and Chen, H.-W., "A combinatorial heuristic algorithm for the three-dimensional packing problem (in Chinese with English abstract)", Journal of Software, 2007, 18(9):2083–2089.

[15] Crainic, T. G., Perboli, G., and Tadei, R., Extreme point-based heuristics for three-dimensional bin packing. INFORMS Journal on Computing, 20(3):368–384.abstract). Journal of Software, 2008, 18(9):2083–2089.

[16] Lodi, A., Martello, S., and Vigo, D., Heuristic algorithms for the threedimensional bin packing problem. European Journal of Operational Research, 2002,141:410–420.

**AUTHOR**

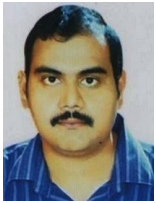 **Drona Pratap Chandu** received Master of Technology in Computer Science and Engineering from Indian Institute of Technology Roorkee, India. He worked as software engineer at Freescale semiconductor India and also he worked as software development engineer at Amazon development centre India recently.